\documentclass[twocolumn]{revtex4}
\usepackage{amssymb,amsmath,amsfonts,bm}
\usepackage{graphicx}
\usepackage{color}

\newlength{\plotwidth}
\begin{document}

\date{\today}

\title{Evolution of inverse cascades and formation of precondensate in\\
  Gross-Pitaevskii turbulence in two dimensions}

\author{Natalia Vladimirova}
\affiliation{
University of New Mexico, Department of Mathematics and Statistics, Albuquerque NM 87131
}

\begin{abstract}
Here we study how coherence appears in a system driven by noise at
small scales. In the wave turbulence modeled by the Gross-Pitaevskii /
nonlinear Schr\"odinger equation, we observe states with correlation
scales less than the system size but much larger than the excitation
scale. We call such state precondensate to distinguish it from
condensate defined as a system-wide coherent state.  Both condensate
and precondensate are characterized by large scale phase coherence and
narrow distribution of amplitudes. When one excites small scales,
precondensate is achieved relatively quickly by an inverse cascade
heating quasi-equilibrium distribution of large-scale modes.  The
transition from the precondensate to the system-wide condensate
requires much longer time. The spectra of precondensate differ from
quasi-equilibrium and are characterized by two bending points, one on
the scale of the average distance between vortex pairs, and the other
on the scale of the distance between vortices in a pair. We suggest
temporal evolution laws for both lengths and use them to predict the
probability of the transition to condensate.
\end{abstract}


\maketitle

%--------------------------------------------------------------

In nonlinear systems, a conserved quantity can be distributed among
large number of degrees of freedom.  Such systems are commonly studied
in spectral space where nonlinear interaction of modes becomes more
apparent.  If the conserved quantity is deposited in a narrow range of
modes, or on a particulate length scale, larger and smaller scales
eventually become excited.  The most notable examples are
redistribution of energy between scales of fluid turbulence and
redistribution of wave action in wave turbulence.  Presence of a
second conserved quantity (enstrophy in two-dimensional fluid
turbulence or energy in wave turbulence) additionally requires
transfer to large scales in the so-called inverse cascade. Unless
infinite space is considered, the inverse cascade is restricted by the
size of the system.  The persistent excitation of small scales can
lead to accumulation of conserved quantity on the scale of the system
--- turbulent formation of condensate.  In two-dimensional fluid
turbulence the condensate appears as a system-wide vortex; in wave
turbulence the condensate is a background state with fast-rotating
phase and uniform intensity.

When separation of scales is large, the condensate can be difficult to
build up.  There is no general recipe on how long and how strongly one
needs to pump the system to observe the condensate. The shape of
evolving spectrum is not known as well.  In the weak wave turbulence
theory~\cite{ZakharovLvovFalkovich1992}, which assumes interaction
local in $k$-space and reduces the description to a kinetic wave
equation, front-like spectra were observed for inverse and direct
cascades in hydrodynamic turbulence~\cite{falsha1991} and for direct
cascades in more general settings~\cite{connew2003}.  In the models
that account for phase interactions of modes, such as Gross-Pitaevskii
(GP) model~\cite{PitaevskiiStringariBook2003}, the spectra can spread
out rapidly with nontrivial shapes, as was shown
in~\cite{dnpz,nazono2006}.  This suggests importance of nonlocal
interactions in the GP system.

The  Gross-Pitaevskii equation, also
known as nonlinear Schr\"odinger equation, is one of the most studied
in modern physics because of its universality. The equation is
applicable to a wide range of phenomena in fluids, solids and plasma,
including non-equilibrium states of cold atoms in Bose-Einstein
condensates~\cite{PitaevskiiStringariBook2003} and propagation of
light in media with the Kerr nonlinearity~\cite{SulemSulem1999}.
In two dimensions, the equation describes evolution of complex wave
envelope $\psi$,
\begin{equation}
  \psi_t=i\nabla^2 \psi + is |\psi|^2\psi,
  \label{NLSE}
\end{equation}
with wave action $N=\langle|\psi|^2\rangle$ being the conserved
quantity in question.  Here, $s$ distinguishes focusing/attractive
($s=+1$) and defocusing/repulsive ($s=-1$) nonlinearity and the
angular brackets denote averaging in space.
%${\cal  N}=\int|\psi|^2d{\mathbf r}$.

When applied to the GP equation, weekly-nonlinear theory predicts
formation of large structures for both focusing and defocusing
nonlinearity~\cite{ZakharovLvovFalkovich1992}.  However, with increase
of nonlinearity, these large structures become unstable if
nonlinearity is focusing~\cite{Bogoliubov1947}.  This suggests that
the condensate can be observed only in the defocusing
case~\cite{dnpz,DyachenkoFalkovich1996}. Accumulation of wave action
in defocusing systems leads to a different kind of coherent structures
--- vortices, i.e. locations with zero amplitude, around which the
phase makes $2\pi$ turn.  As shown in
Refs.~\cite{nazono2006,nowsch2012,schnow2012}, decrease in number of
vortices leads to formation of condensate.
 
In this paper, we study the evolution of turbulence in GP model during
persistent excitation of small scales.  Our goals are (i) to explore
possibility of appearance of local order on scales smaller than the
domain size; (ii) establish a connection between time-dependent
spectra and phase coherence of the system, in particular evolution of
vortices; (iii) study the effect of system size on its evolution and 
make qualitative prediction on probability of formation of system-wide
condensate in domains of finite size.

%-------------------------------------

We stress that the key focus of this study is turbulence evolution.
Our earlier work~\cite{falvla2015} was devoted to the fluxes of direct
and inverse cascades in a steady state of GP system stabilized by
large-scale friction.  While we have observed some mid-range
distortion of the spectra, which was independent of the domain size
and similar to the described below, the large scale modes were
suppressed by friction.  It is those modes that influence the
mid-range modes via non-local interactions, making the distortion the
feature of steady spectra, in the way the obtained flux law was
specific to steady system.  Naturally, a steady setup cannot inform
one on the timescale of establishment of the condensate, while here we
propose a quantitative estimate of the time of formation of the
condensate under constant pumping rate. 

\begin{figure}
\begin{center}
\includegraphics[width=1.02\plotwidth]{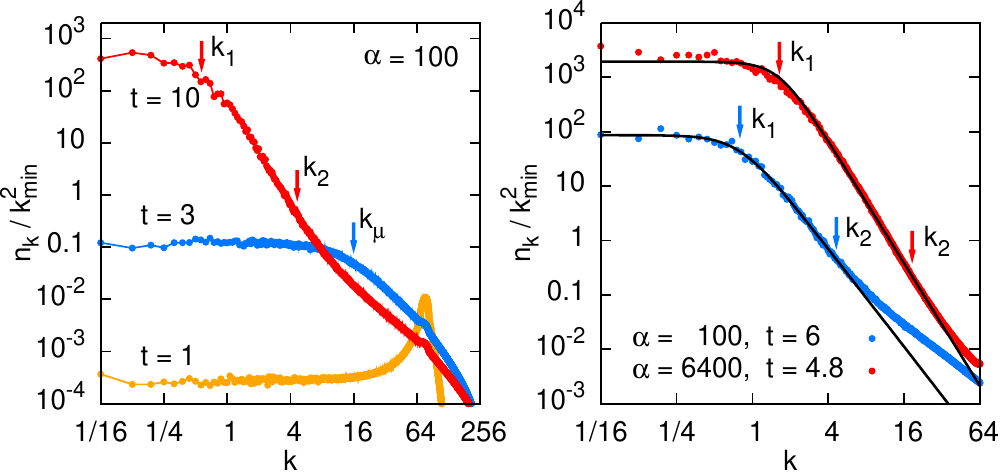}
\caption{
  Left: Typical spectra of wave action at very early, early, and late
  stages of evolution.  Right: Spectra at late times fitted by
  Eq.~\eqref{eq:late_spectra_fit}.  Here, $k_\mu$ marks the bending
  point in early spectra, while $k_1$ and $k_2$ are
  bending points in late spectra; $k_{\min}=2\pi/L$, where $L$ is the 
  size of the system.}
\label{fig:evol}
\end{center}
\end{figure}

Similarly to~\cite{falvla2015},
we numerically solve Eq.~\eqref{NLSE} with focusing nonlinearity as
described in Supplement~\ref{supl:numerics}.  The wave action is
deposited at the rate $\alpha$ in a ring of wave numbers at $k\approx
k_p$, and accumulates in the system at the rate $\tilde{\alpha} \approx
0.92 \alpha$.  Initially, the spectrum of wave action is empty, $n_k
\equiv |\psi_k|^2=0$.  The first excited modes appear in the pumping
ring. Our intuition might tell us that modes with close $\mathbf k$
interact more effectively, resulting in gradual widening of the
spectrum beyond the pumping ring.  Apparently this is not the
case. Already after the time period comparable with the nonlinear
interaction time, we observe uniform distribution of $n_k$ for $k <
k_p$, as well as for $k>k_p$.  The spectrum at $k < k_p$ remains flat,
with $n_k$ growing in time; the spectrum at $k>k_p$ in more complex as
it is affected by damping.  We observe the scaling $n_k \propto
\alpha$ for the forced modes and scaling $n_k \propto \alpha^3$ for
the non-forced modes. The second scaling follows from the first one
and from cubic nonlinearity.  The flat shape of $n_k$ most likely is
the consequence of circular arrangement of forced modes.  The
simultaneous growth of all modes illustrates the importance of
nonlocal interactions already at the beginning of evolution. Indeed,
the plateau that extends from $k=0$ to the forcing ring is a
characteristic of very early spectra, as seen in Fig.~\ref{fig:evol}.

With time, the peak at the forcing becomes smaller, the height of
plateau rises, and a section of sloped spectrum develops between the
plateau and the forced modes. This shape of the spectrum can be
described by time-dependent energy-action equipartition:
\begin{equation}
n_k = \frac{T(t)}{k^2_\mu(t) + k^2}\,,
\label{eq:thermal_nk}
\end{equation}
where $T$ and $\mu\equiv k^2_\mu$ can be interpreted as temperature
and chemical potential. The temperature controls the height of the
sloped part of the spectra, $n_k\approx T/k^2$, while $k_\mu$
corresponds to the bending point at the end of the plateau.

\begin{figure}
\begin{center}
\includegraphics[width=1.0\plotwidth]{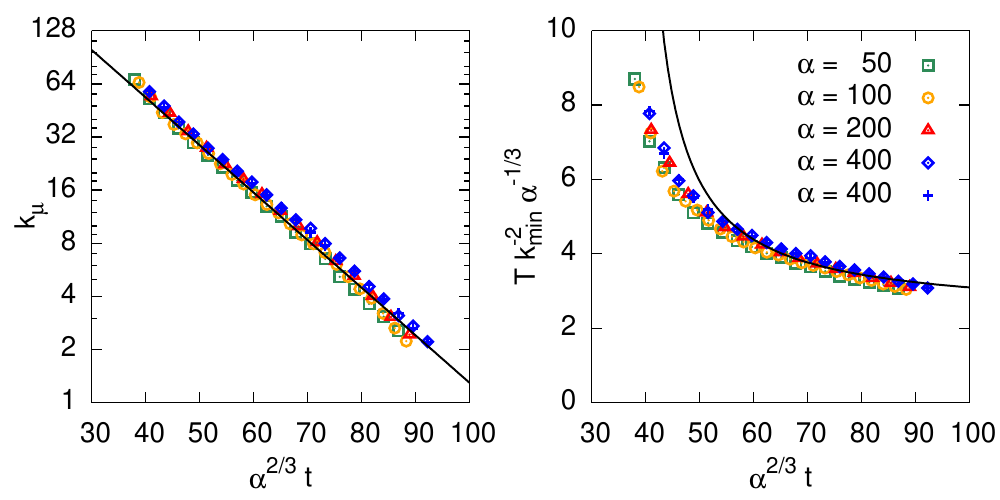}
\caption{
  Chemical potential and temperature at
  early times for different $\alpha$.  All points, except
  crosses, are data from simulations with $L=8\pi$; crosses are data from
  simulations with $L=32\pi$.  The lines show dependence
  $k_\mu = C \exp(-c \alpha^{2/3} t)$ and
  $T$ given by Eq.~\eqref{Teq} with $C=640$ and $c=0.062$.
}
\label{fig:muT}
\end{center}
\end{figure}

Both $T$ and $k_\mu$ decrease with time, as shown in
Fig.~\ref{fig:muT}.  The fit by Eq.~\eqref{eq:thermal_nk} can be
applied to the data only when $k_\mu < k_p$; yet, the very early rise of
the flat spectrum can be seen as the same process.  Initially,
chemical potential is so large that $k_\mu(t)>k_p$ and most of the
waves at $k<k_p$ appear in the state of action equipartition. Filling
the system with waves, we decrease the chemical potential; after
$k_\mu$ decreases below $k_p$ we start seeing the part of energy
equipartition $n_k\propto k^{-2}$ simultaneously with the rise of the
plateau.

As shown in Fig.~\ref{fig:muT}, the data from simulations at different
pumping rates, $\alpha$, and in domains in different sizes collapse
onto a single curve when rescaled with $\alpha$.  The decay of $k_\mu$ is
exponential, while $T$ approaches an asymptote.  The exponential
decay of $k_\mu$ follows from the linear growth of the wave action,
$\int n_k d\mathbf{k} \approx T\ln (k_p/k_\mu) \simeq \tilde\alpha t$,
under assumption that the temperature must eventually saturate.  Then,
assuming dependence $k_\mu = C \exp(-c\alpha^{2/3}t)$, suggested by
data, one can find the temperature in the limit of $k_\mu \ll k_p$,
\begin{equation}
  \frac{T}{k^2_{\min}} = (2\pi)^{-1}
    \frac{\tilde{\alpha}t - N_p}{c\alpha^{2/3}t - \ln(A/k_p)},
\label{Teq}
\end{equation}
where $k_{\min}=2\pi/L$ in a system of size $L$, and $N_p$ is the
number of waves at $k>k_p$; in our simulations $N_p \approx
c^{-1}\alpha^{1/3}$.  The dependence explains collapse of data in
coordinates $(\alpha^{2/3}t, \, T \alpha^{-1/3})$ observed in
Fig.~\ref{fig:muT}.

The scaling of temperature with
$\alpha$ is consistent with weakly nonlinear theory.  We expect that,
if nonlinearity is weak, the flux is cubic in wave numbers
$n_k$~\cite{ZakharovLvovFalkovich1992}, so that $T\propto
\alpha^{1/3}$ and $\ln k_\mu\propto \alpha^{2/3}t$, which is indeed
seen in Fig.~\ref{fig:muT}.  In general, the scaling for temperature
and conservation of wave action lead to the scaling of time
with~$\alpha$,
\[
\dot{N} \sim \dot{T} \sim \alpha \; \Rightarrow \;
T/t \sim \alpha \; \Rightarrow \;
t \sim \alpha^{1/3} \alpha^{-1} \sim \alpha^{-2/3}.
\]

One might find the decrease of the temperature with time
counter-intuitive. We think it can be interpreted again in terms of
nonlocal interaction: to carry the same flux through a longer
spectrum one needs smaller amplitude.  In other words, nonlocal
transfer of wave action through a given $k$ is determined by both an
amplitude and an extent of the interval. When the interval expands
towards lower $k$ and acquires higher $n_k$ at low $k$, the transfer
becomes more effective, and the magnitude decreases.

%------------------------------------------------------
%\section{Late evolution and formation of pre-condensate}
%\label{late}
%------------------------------------------------------

%\subsection{Coherence in amplitude}

As time passes and wave action accumulates, the system transitions to
a different regime where the spectra have two bending points and the
fit by Eq.~\eqref{eq:thermal_nk} no longer applies (see
Fig.~\ref{fig:evol}).  A similar transition occurs in systems with
focusing nonlinearity, as shown in Supplement~\ref{supl:focusing}. The
transition time $t^* \approx 90 \alpha^{-2/3}$ and corresponding
$k_\mu \approx 2$ are surprisingly universal.  Moreover, as we show
below, the scaling $\alpha^{2/3} t$ well describes evolution in the
new nonlinear regime, even though this scaling was obtained under
assumption of week nonlinearity.  This is somewhat surprising.

\begin{figure}
\begin{center}
\includegraphics[width=\plotwidth]{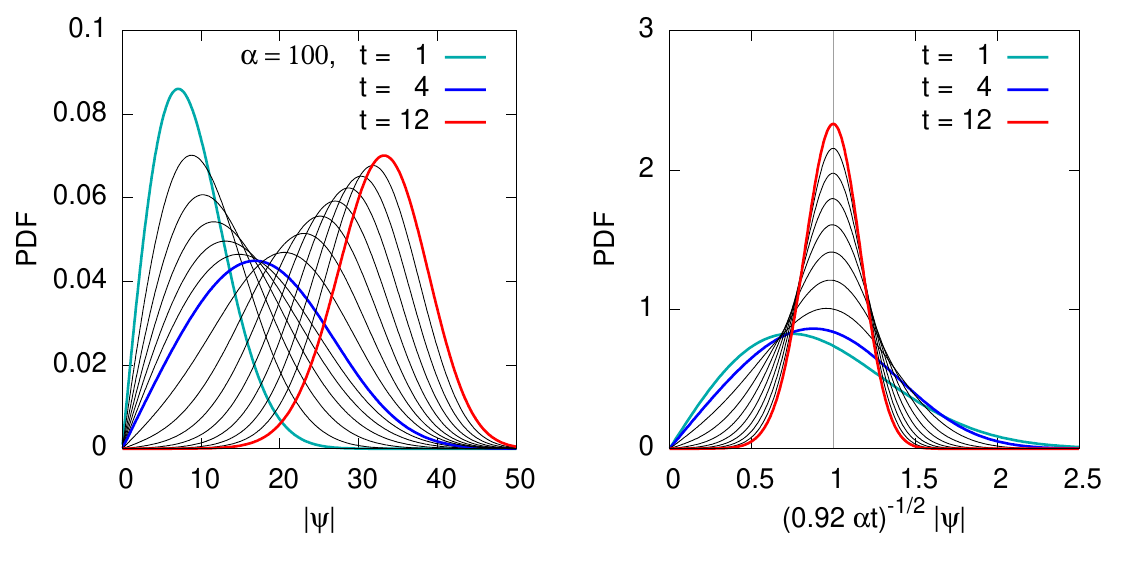}
\caption{
  PDF of $|\psi|$ in simulation units (left) and
  in unit rescaled rescaled with $N = \tilde{\alpha} t = 0.92\,\alpha t$
  (right).
}
\label{fig:psipdf}
\end{center}
\end{figure}

Even more dramatically than in spectra, the transition to the new
regime is seen in the probability density function for $|\psi|$, shown
in Fig.~\ref{fig:psipdf}.  Here, we follow the evolution of
distribution of amplitudes with respect to time-dependent average,
$\chi = |\psi|/|\psi|_{\rm rms} = N^{-\frac{1}{2}} |\psi|$.  At the
early stage, the distributions of real and imaginary parts of $\psi$
are Gaussian with zero average, so that the distribution of magnitude
has the form ${\cal P}(\chi) = 2\chi e^{-\chi^2}$; at this stage the
standard deviation for $|\psi|$ widens with time, $\sigma =
\frac{1}{2} N^{1/2}$.

In contrast, in the new regime the distribution narrows and shifts
toward higher amplitudes. The maximum is located at $|\psi|= N^{1/2}$,
while the overall shape closely resembles a Gaussian, $\ln{\cal
  P}(|\psi|)\propto -(|\psi|-N^{1/2})^2/\sigma^2(t)$.  The probability
of small fluctuations, $|\psi| \ll N^{1/2}$, is determined by vortices
(see Supplement~\ref{supl:vortices} for more detail).
Figure~\ref{fig:evol_pdf} shows the growth of $\langle|\psi|\rangle$,
which scales as $t^{1/2}$ during both early and later stages; it also
shows non-monotonic time-dependence for $\sigma$. The time when the
distribution is the widest is easily detectable, $t^* \approx 90
\alpha^{-2/3}$. We use this time as the definition for transition
between the earlier and later regimes.

\begin{figure}
\begin{center}
  \includegraphics[width=\plotwidth]{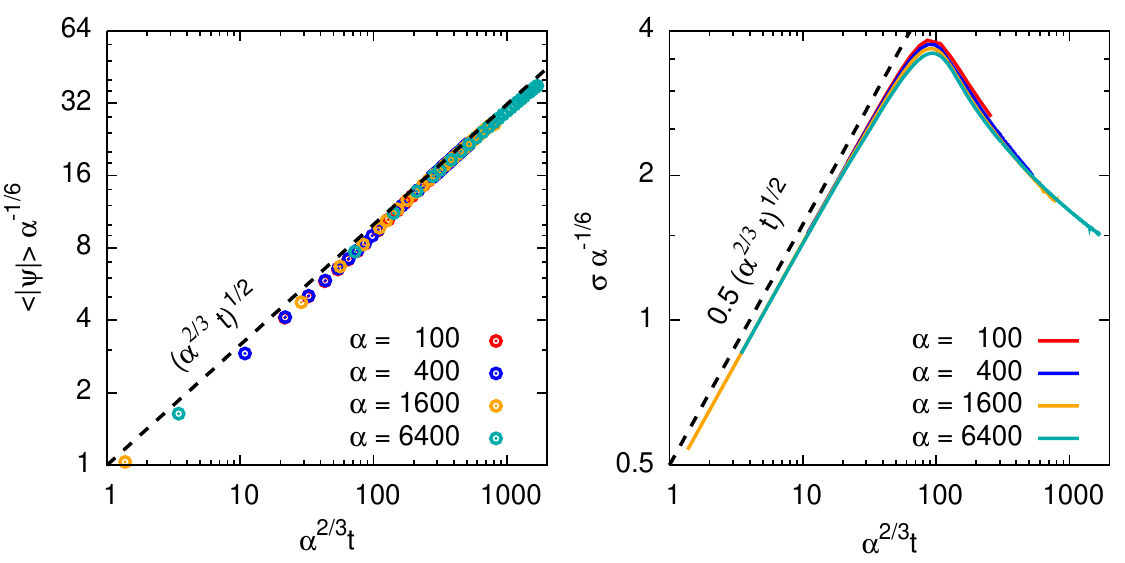}
\caption{
  Average (left) and standard deviation (right) for distribution of $|\psi|$
  as function of time, in units scaled with $\alpha$.
}
\label{fig:evol_pdf}
\end{center}
\end{figure}

%----------------------------------------------------------------
%\subsection{Phase coherence and vortices}
%----------------------------------------------------------------

The narrow distribution of $|\psi|$ is a prominent feature of
system-wide condensates, where most of the wave action resides in the
single mode, $k=0$, while other waves add small distortion to
condensate's background. In the case considered here, the background
is formed by multiple modes, so we refer to the state at $t>t^*$ as
``precondensate'', as opposed to system-wide condensate.

Another feature of condensates is the phase coherence.  In a system
with a system-wide condensate there is no vortices and the phase
across the domain only slightly deviates from the phase of zeroth
mode.  Precondensate at its later stages can have most of the wave
action absorbed in $k=0$ mode, yet only partial phase coherence
because of the presence of vortices.  In such cases the scale of phase
coherence is the typical distance between
vortices~\cite{nazono2006,nowsch2012,schnow2012}.

Our simulations show that $t^*$ corresponds to the time when distinct
vortices start to appear.  At $t<t^*$ the probability of near-zero
$|\psi|$ is high, the phase correlation length is short, and formal
detection of vortices returns vortex locations all over the
computational grid. If vortex is a ``hole'' in the background
amplitude, to have vortices we need to have a non-zero background.  At
$t \approx t^*$ such background begins to form.

During time interval $t^* \lesssim t \lesssim 2 t^*$, the distance
between detected vortices is still of the order of grid resolution,
but the number of vortices drops sharply.  At $t \gtrsim 2 t^*$, the
vortices can be located by visual inspection of phase field; their
number decreases in time, but much slower.  One can think of the state
at $t<t^*$ as containing no distinct vortices, time interval $t^*< t <
2t^*$ as the stage of vortex formation, and $t>2t^*$ as the stage of
vortex annihilation.

\begin{figure}
%\begin{center}
%\includegraphics[width=1.0\plotwidth]{FIGS5/four_corners.png}
\includegraphics[width=1.0\plotwidth]{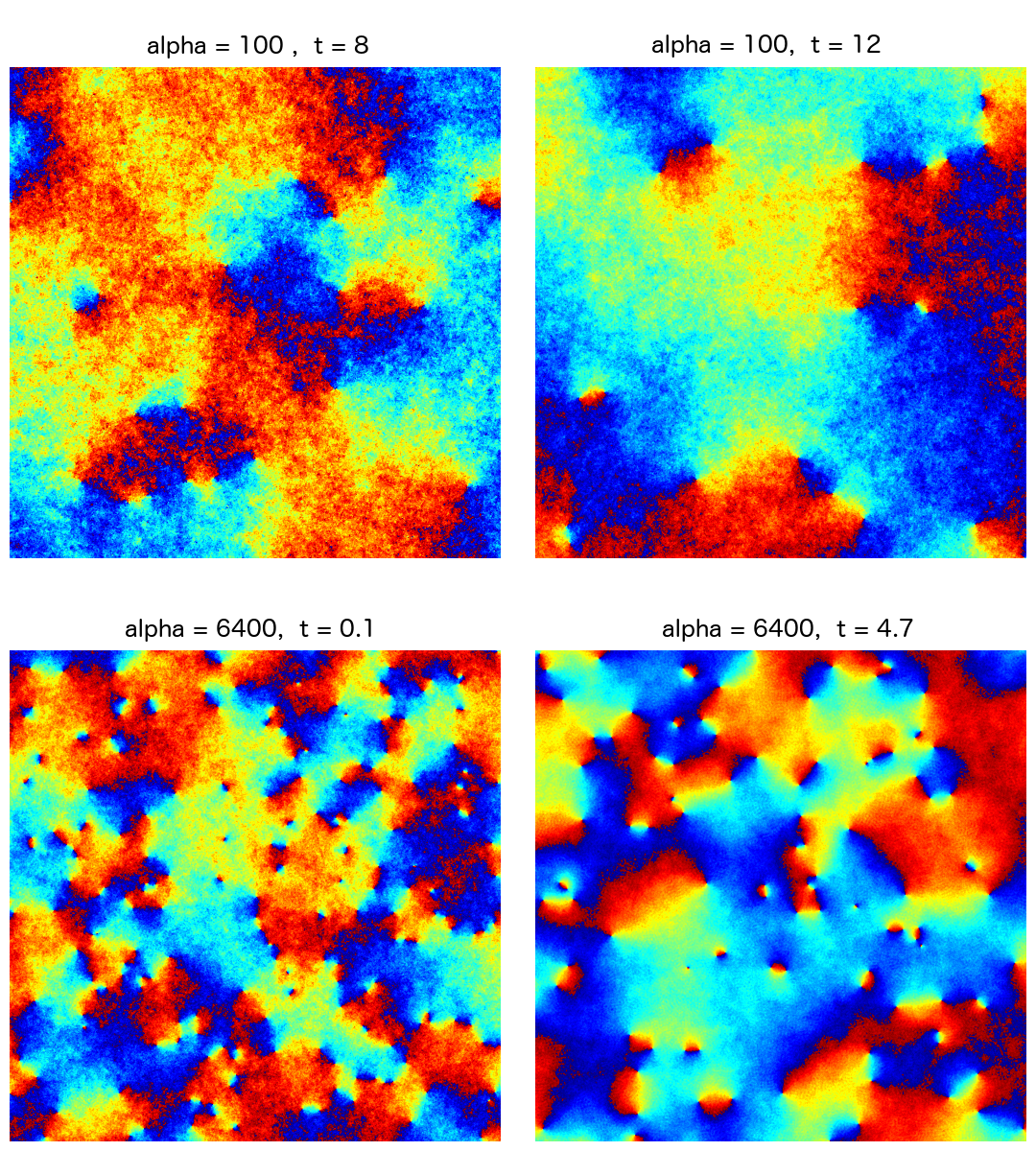}
\caption{
  The phase in a fraction of the domain, $L/8 \times L/8$, from
  simulations with $L=32\pi$.  The images illustrate that (i) the
  typical distance between vortex pairs, $d_1$, increases in time,
  (ii) the typical scale of the vortex pair, $d_2$ remains constant,
  and (iii) the distance between the vortices in a pair, $d_2$, is
  smaller for larger $\alpha$.  }
\label{fig:vortices}
%\end{center}
\end{figure}

Figure~\ref{fig:vortices} shows snapshots of phase for two pumping
rates at two times.  Notice that the system with $\alpha=6400$ and
$t=4.7$ has smoother phase than the one with $\alpha=100$ and
$t=8$; this is because the transition timescale is shorter for
stronger pumping, $t^*=0.26$ versus $t^*=4.2$.  Also notice that
vortices form pairs; and that the typical distance between vortices in
a pair remain constant on the course of evolution, while the number of
pairs decreases. And finally, notice that the system with higher
pumping rate has more vortex pairs and shorter distance between
vortices in a typical pair.

To quantify these observations, we have implemented diagnostics of
vortices and vortex pairs, described in
Supplement~\ref{supl:vortices}.  If we denote the number of vortices
of the same sign (half of total number of vortices) as $n_{\rm vort}$,
then the typical distance between isolated vortices or vortex pairs is
$d_1 = Ln^{-1/2}_{\rm vort}$.  The typical distance between vortices
in a pair, $d_2$, is estimated from the distribution of distances
$d_2^{(i)}$ of individual pairs.

We found that the number of vortices scales with $\alpha^{2/3}t$ and
decreases with time. The time range is too short to suggest a
functional dependence; while both a power law and a logarithmic
dependence are possible, for interpolation purposed we adopted the
power law.  The length of a vortex pair depends on the
pumping rate, rather than time, which is surprising and deserves further
investigation, as discussed in Supplement~\ref{supl:vortices}.

%--------------------------------------------------------------
%\subsection{Spectra of wave action}
%-----------------------------------------------------

Next, we connect statistics of vortices to the evolution of spectra,
$n_k(t)$.  In the precondensate regime, $t>t^*$, the spectra have two
bending points, $k_1$ and $k_2$, as seen in Fig.~\ref{fig:evol}.  An
equipartition shelf at small $k$ meets a slope steeper than $k^{-2}$
at $k=k_1$, this slope transitions to a slope close to $k^{-2}$ at
$k=k_2$.  We fit the spectra in the range $[k_{\min}, k_2]$ using the
following function,
\begin{equation}
n_k = \frac{A}{1 + (k/k_1)^p}.
\label{eq:late_spectra_fit}
\end{equation}
Here, $p$ is some power and $A$ is the height of the equipartition
shelf.  When $p=2$, $k_1 = k_\mu$, and $A=T/k^2_\mu$, the fit reduces
to Eq.~\eqref{eq:thermal_nk}.

\begin{figure}
%\begin{center}
%\includegraphics[width=1.0\plotwidth]{FIGS5/evol4_fit.pdf}
\includegraphics[width=1.0\plotwidth]{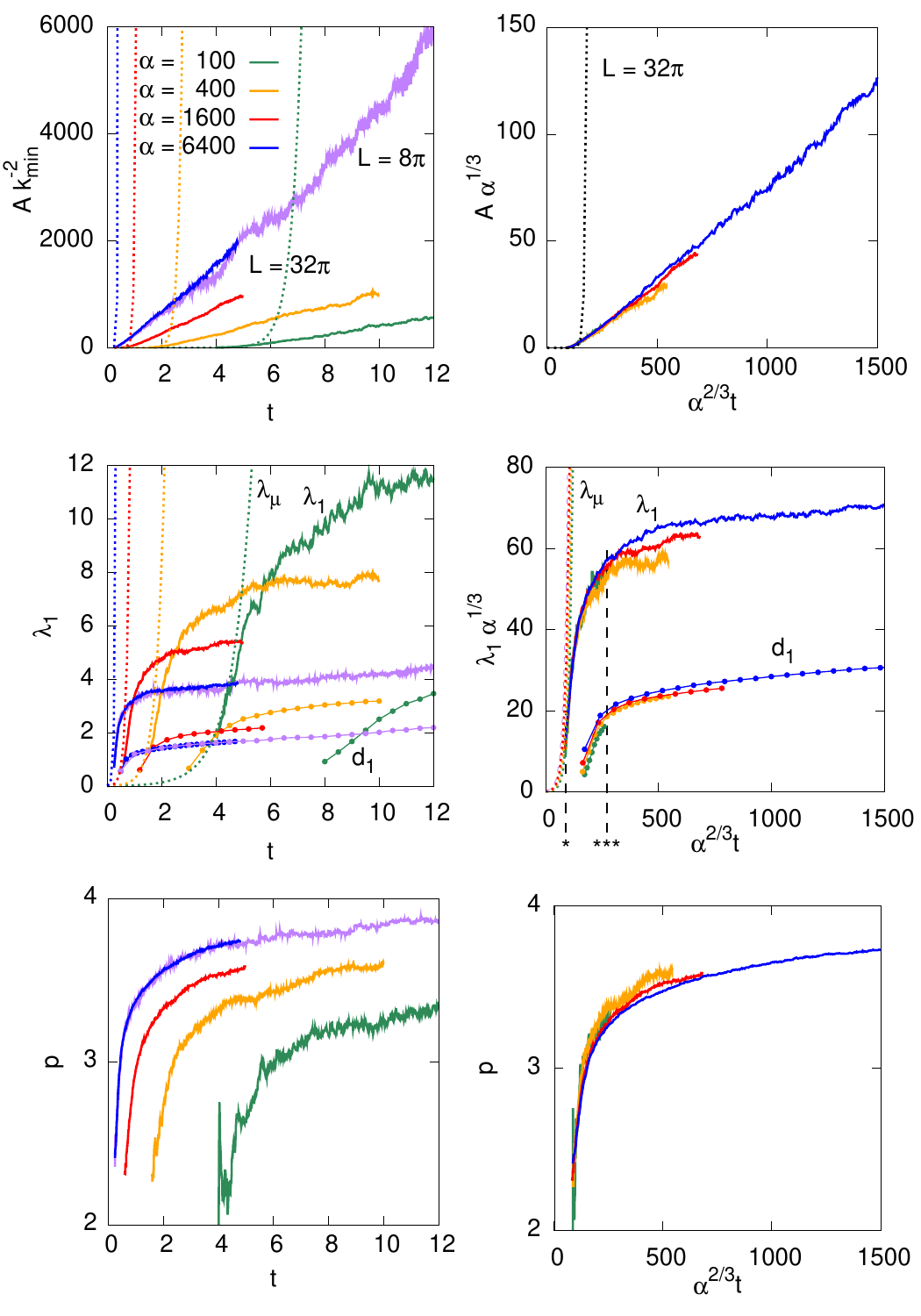}
\caption{
  Parameters in late evolution spectra, $A$, $\lambda_1 = 2\pi/k_1$,
  and $p$, as function of time in simulation and rescaled units.
  Purple line corresponds to $\alpha = 6400$, $L=8\pi$; all other data
  are from $L=32\pi$ simulations.  Dotted lines show the
  extrapolations from weakly-nonlinear regime, $A=T/k^2_{\mu}$ and
  $\lambda_\mu = 2\pi/k_{\mu}$.  In the middle row, the typical
  distance between vortices, $d_1 = L n^{-1/2}_{\rm vort}$, is
  shown with connected points, and dashed vertical lines
  indicate the times $t^*$ and $3t^*$.
}
\label{fig:evol_spectra}
%\end{center}
\end{figure}

Figure \ref{fig:evol_spectra} shows how parameters of spectra in
Eq.~\eqref{eq:late_spectra_fit} change with time.  After transition,
the height of the shelf rises linearly with time, as $A \propto
\alpha^{1/3} t$, in contrast to early evolution, when the height of
the shelf, $T/k^2_\mu$ grows exponentially (due to exponential decay of
$k_\mu$).  The scale associated with the first bending point in the
spectra, $\lambda_1 = 2\pi/k_1$, initially grows rapidly, yet not as
fast as exponentially increasing $\lambda_\mu = 2\pi/k_\mu$.  At the time
$t \sim 3 t^*$ the growth of $\lambda_1$ slows down.  The scale
$\lambda_1$ appears to be proportional to the distance between vortex
pairs, $\lambda_1 \sim 2.5 d_1$, so we conclude that the first bending
point in the spectra marks the scale of phase coherence, or the scale
of patches of precondensate.  The third parameter, power $p$,
describes the slope of the spectrum after the plateau, in $k_1< k<
k_2$ range. This slope steepens with time from $p=2$ in
thermal equilibrium regime to possibly $p=4$ in long-run evolution.

If $k_1$ corresponds to the distance between vortex pairs $d_1$, the
natural question arises --- what scale corresponds to $k_2$?  We
notice that the second bending point of the spectrum does not move on
the course of evolution; however, it shifts to the smaller scales as
$\alpha$ increases.  Recall that we made the same observation about
the typical distance between vortices in a pair, $d_2$. And indeed,
the corresponding wave number $k_2 = {\alpha}^{1/3}$ is located
approximately at the second bending point of the spectra (see
Supplement~\ref{supl:vortices} for detail).  Thus, the data suggest
that $\lambda_2 = d_2 = 2\pi{\alpha}^{-1/3}$ is the typical distance
between vortices in a vortex pair.

The emergence of second inflection point and deviation from thermal
equilibrium spectra can be interpreted as an internal ``bottleneck''
effect.  The pile-up occurs at wave numbers where nonlinearity is
getting substantial.  We have observed a similar pile-up in
simulations stabilized by low-$k$ friction~\cite{falvla2015}, where,
regardless of the domain size, stronger nonlinearity leads to more
pile-up, while pumping at lower rate reduces piling-up and extends the
universal part of the spectrum.

Until now we have studied the evolution of wave turbulence before it
gets affected by the size of the domain.  Now we are interested in the
transition from precondensate to a system-wide condensate.  We need
relatively long simulations in relatively small boxes, so we can watch
all vortices disappear.  We expect this to happen when the typical
distance between vortices, $d_1$, exceeds the domain size.

We found that the number of vortices in domains with sizes $L=2\pi$,
$\pi$, and $\pi/2$, follow the same dynamics as our large-scale
simulations, $L=32\pi$, provided that the evolution in small boxes is
interpreted in the statistical sense. (We have considered ensembles of
multiple realizations for each combination of parameters; see
Supplement~\ref{supl:condensate} for detail.)  The chances of
transition to condensate is are much higher when $d_1$ exceeds the
domain size during vortex formation stage; during vortex annihilation
stage precondensate slows down the annihilation of vortices. For
creation of system-wise condensate slow pumping rates are favorable,
since $d_1$ is an increasing function of $\alpha^{2/3}t$. In general,
one can predict the typical time of transition to condensate by
solving $d_1(\alpha^{2/3}t) = L$.  This statement is not obvious,
since one could expect the limited size of the system to have an
additional effect.

Once the condensate has established, the spectra for over-condensate
fluctuations are expected to have slope $n_k\propto k^{-2}$
\cite{dnpz,vlader2012}.  Unfortunately, we could not detect the
transition from $k^{-p}$-spectrum for a precondensate to the
$k^{-2}$-spectrum of over-condensate fluctuations.  This is because to
resolve precondensate spectra, we need many modes and large domains,
while slow annihilation of vortices requires long simulation times.
All simulation, where we could achieve transition to system-wide
condensate, are done in small boxes. In these simulations, the spectra
never have a chance to develop slopes with $p>2$.  Instead, they
transition from the thermal equilibrium spectrum with $p=2$ directly
to the spectrum with $p=2$ for over-condensate fluctuations.

%-----------------------------------------------------
\subsection*{Conclusion}
%-----------------------------------------------------

In this work, we have used the model of Gross-Pitaevskii / nonlinear
Schr\"odinger equation to study evolution of wave turbulence excited
by small-scale forcing.  While the wave action accumulates in a system
at a constant rate, there is a time $t^*$ that marks transition from
weakly nonlinear to substantially nonlinear regime (when focusing case
and defocusing case start to deviate, as shown in
Supplement~\ref{supl:focusing}). At $t<t^*$ spectra of $n_k$ have the form
of time-dependent energy-action equipartition, while the distribution
of $|\psi|$ widens with time. At $t>t^*$ the distribution of $|\psi|$
in the defocusing case concentrates near rising background
(precondensate), while spatial locations with near-zero $|\psi|$
become sparse and develop vortex structure.  The typical distance
between vortex pairs and the typical distance between vortices in a
pair correspond to two bending points in spectra of wave action.

Evolution of vortex density in a large domain well describes
probability of developing a system-wide condensate in a small domains.
The condensate is more likely to appear if the number of vortex pairs
is expected to drop below 1 during vortex generation stage, $t^*
\lesssim t \lesssim 3t^*$. Later, at $t \gtrsim 3t^*$, strong
precondensate prevents vortex interaction, and vortex annihilation
slows down. The rescaling between nondimensional units and physical
units and estimate for the transition time, $t^*$, in physical units
in shown in Supplement~\ref{supl:units}.

\subsection*{Acknowledgement}

I thank G.~Falkovich for encouragement and discussions, and for
reading the draft of the paper.  The work is supported by NSF grant
no. DMS-1412140. Simulations are performed at Texas Advanced Computing
Center (TACC) using Extreme Science and Engineering Discovery
Environment (XSEDE), supported by NSF grant no. ACI-1053575.

%-----------------------------------------------------
\section*{Supplemental materials}

%-----------------------------------------------------
\subsection{Numerical setup}
\label{supl:numerics}
%-----------------------------------------------------

Our setup is almost identical
to~\cite{falvla2015} where we studied the inverse cascade stabilized by
large-scale friction, with the exception that now the friction is
turned off. 

The wave action is deposited at the rate $\alpha$ in a
ring of wave numbers, $k \in [k_l,k_r]$.  Some fraction of it is lost
to small-scale damping, applied at $k> k_d \approx 3 k_r$, the rest
accumulates in the system at the rate $\dot{N} = \tilde{\alpha}$.  The
forcing and damping are represented in the right hand side of the
equation,
\begin{equation}
   i\psi_t + \nabla^2\psi + s |\psi|^2\psi = i\hat{f_k} \psi + i\hat{g_k}.
  \label{NLSEclean}
\end{equation}
Forcing and damping are both applied in spectral space. The forcing is
additive, $g_k = |g_k| e^{i\phi_k}$, with random phases $\phi_k$ and
amplitudes $|g_k| \propto \sqrt{(k^2 - k_l^2)(k^2_r - k^2)}$, while
the damping is multiplicative, $f_k = - \beta (k/k_d)^4(k/k_d - 1)^2$.
Equation~\eqref{NLSEclean} is solved using a standard split-step
method~\cite{dnpz} modified to be 4th-order accurate in time. 

Our computational domains are square, $L\times L$, with periodic
boundary conditions, so that the lowest wave number is determined by
the domain size, $k_{\min} = 2\pi/L$.  The highest wave number is the
same in all simulations, $k_{\max} = \pi/\Delta x = 512$, as well as
the following parameters, $k_l = 68$, $k_r = 84$, $k_d=256$, and
$\beta=400$.  This choice of parameters gives 8\% loss of wave action
in most of simulations, $\tilde{\alpha} = 0.92 \alpha$.  We model
systems with different strengths of forcing, $\alpha = 100$, 400,
1600, and 6400, and of different sizes, up to $L=32\pi$.
Note that our main results are scaled with $\alpha$ and $k_{\min}$, so
that the forcing length scale is the only fixed parameter in our study.
This restriction can be relaxed by rescaling of units described
in Supplement~\ref{supl:units}.

As a remark on the size of the simulation, we emphasise that major
results reported in this paper --- the pile-up of wave action at low
$k$ and the formation of spectra with two bending points --- are not
effected by a finite domain size.  Most results are obtained in
domains with $L=32 \pi$; yet when we repeated some of simulations in
domains $L=8\pi$, we observed essentially the same behavior, see for
example the curves for $\alpha = 6400$ in Fig.~\ref{fig:evol_spectra}.
The largest of the discussed length scales is $\lambda_1 \sim 10$ (at
the end of the run with $\alpha=100$), which is still small compared
to $L=32\pi \approx 100$. The smallest number of vortex pairs used in
vortex statistics is 840, also at the end of the run with
$\alpha=100$; this number is large enough to ignore the effects of
domain size.

%-----------------------------------------------------
\subsection{Focusing case}
\label{supl:focusing}
%-----------------------------------------------------

\begin{figure}
\begin{center}
\includegraphics[width=\plotwidth]{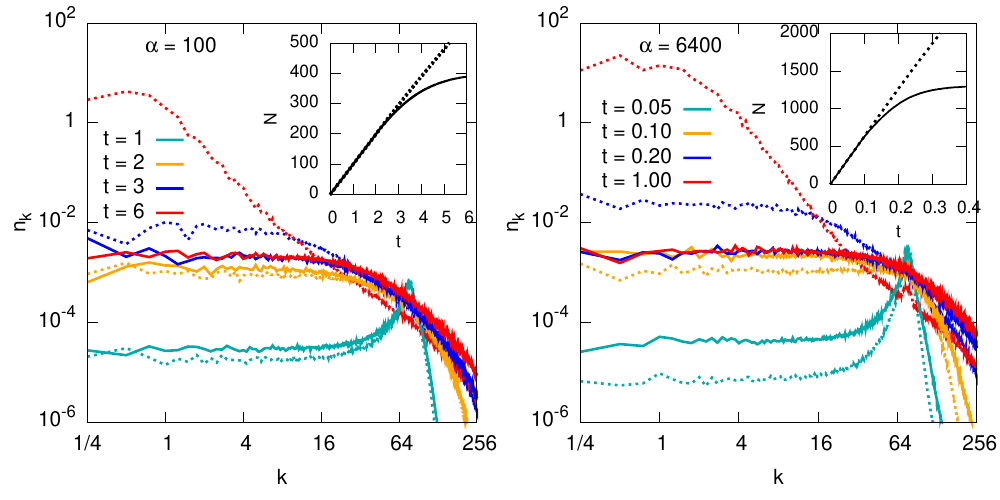}
\caption{
 Comparison of focusing spectra (solid lines) and defocusing spectra
 (dashed lines) at different times of system evolution for
 $\alpha=100$ (left) and $\alpha=6400$ (right).  Inserts show the
 total number of waves.
}
\label{fig:sp_f1}
\end{center}
\end{figure}

Weakly nonlinear theory does not distinguish positive and negative
nonlinearity. So, at very early times, the spectra with focusing and
defocusing nonlinearities are expected to evolve in the same way. It
turns out that this similarity lasts almost to the end of weakly
nonlinear regime, $t\lesssim t^*$. Figure~\ref{fig:sp_f1} shows the
spectra of focusing and defocusing systems for $\alpha=100$
($t^*\approx 4.2$) and for $\alpha=6400$ ($t^*\approx 0.26$). At the
very early times, the spectra look qualitatively the same, except that
the focusing nonlinearity is more effective in populating low-$k$
modes, especially at higher $\alpha$ --- possibly because of nonlinear
shift of frequency and higher effective nonlinearity parameter, $(k^2
\pm N)/k^2$.  Since in the weak turbulence approximation the evolution
of focusing and defocusing systems must be exactly the same, this small
difference in the spectra is already an effect of nonlinearity.

Approximately at the time when defocusing spectra start to deviate
from the thermal equilibrium form, focusing spectra stabilize at an
equilibrium. This is also seen in the total number of waves (inserts
in Fig.~\ref{fig:sp_f1}). The stabilization of $N(t)$ at constant
pumping is the sign of enhanced loss of wave action due to
collapses. Indeed, at $t\approx t^*$ both systems start to develop
coherent patches of precondensate.  In the focusing case, coherent
patches turn into collapses; this process transfer wave action to high
$k$, where it gets consumed by damping.  The stable level of wave
action can be estimated as $N^* \approx \alpha t^* \propto
\alpha^{1/3}$.  By the order of magnitude this is seen in simulations,
however the functional dependence appears to be more complex.

The last observation suggests that it might be possible to build
condensate in a focusing system, if the size of domain is so small
that condensate is formed before the total number of waves reaches
critical, $N < 11.7/L^2$.  In our setup, however, this would require
long simulation times at small pumping rates, $\alpha \propto N^3
\propto L^{-6}$ and $t\sim N \alpha^{-1} \sim L^4$.

%-----------------------------------------------------
\subsection{Vortices: diagnostics, evolution, and relation 
to probability of small amplitudes and to spectra}
\label{supl:vortices}
%-----------------------------------------------------

\begin{figure}
%\begin{center}
\includegraphics[width=1.0\plotwidth]{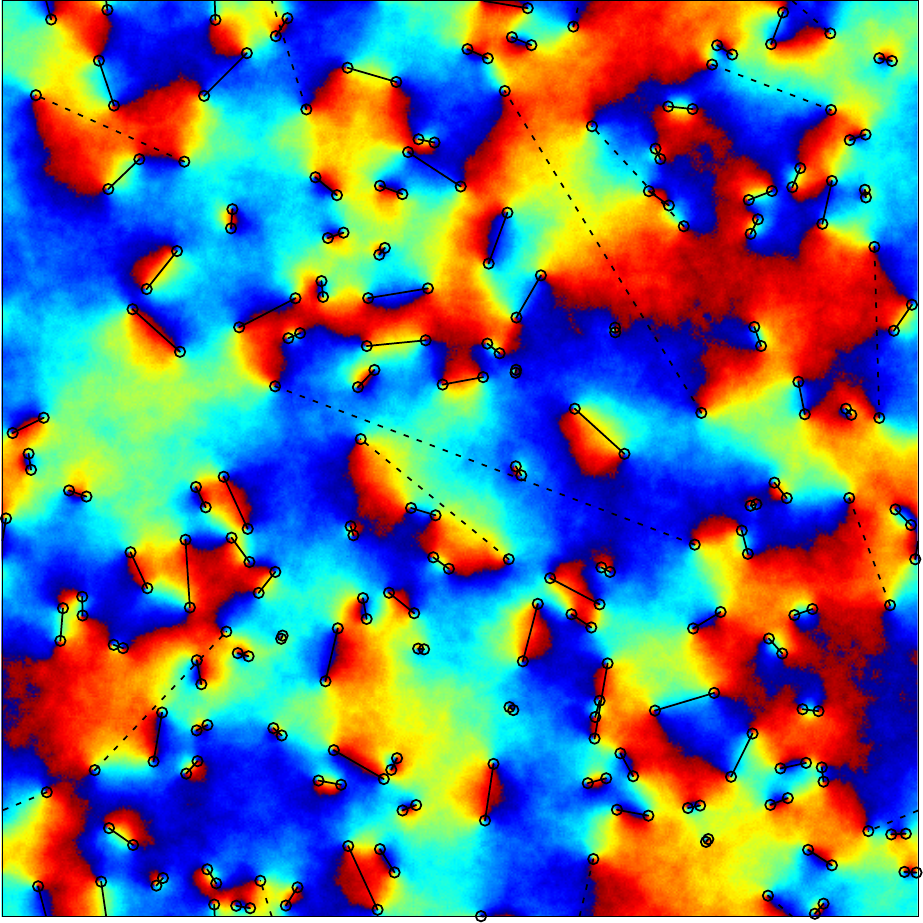}
\caption{
  Diagnostics of vortex pairs. The image shows phase in the system
  with $\alpha = 6400$ and $L=8\pi$ at $t=12$.  The pairs with
  distance between vortices exceeding $2d_1$, where $d_1$ is the
  typical distance between vortex pairs, are marked with dashed
  lines.}
\label{fig:pairs}
%\end{center}
\end{figure}

To find the location of vortices we use a method based on the vortex
definition.  Starting with the phase on a computational grid,
$\phi^j_i$, we compute circulation of phase along the perimeter of
each computational cell,
\begin{eqnarray*}
\delta \phi = &&
       \left[ \phi^j_{i+1} - \phi^j_{i} \right] +
       \left[ \phi^{j+1}_{i+1} - \phi^j_{i+1} \right] +\\
     &&  \left[ \phi^{j+1}_{i} - \phi^{j+1}_{i+1} \right] +
       \left[ \phi^j_{i} - \phi^{j+1}_{i} \right].
\end{eqnarray*}
We restrict each expression in square brackets not to exceed $\pi$ in
absolute value, by adding or subtracting $2\pi$ as necessary.  The
cells with nonzero $\delta \phi$ are recorded as vortices. (We have
observed only vortices with single charge, $\delta \phi = \pm 2\pi$.)

To find vortex pairs, we compute matrix of distances between positive
and negative vortices. Two vortices with the shortest distance are
assigned into a pair and excluded from the list. Then, the pair with
the shortest distance is found again from the reduced matrix, and the
process is repeated until all vortices are assigned into pairs. This
might not be an optimal algorithm, say in comparison with minimizing
sum of distances over all possible pair assignments, but it is easy
to implement and fast to execute. A side effect of this algorithm is a
small number of distant vortices formally assigned into pairs; this
happens at the end of assignment procedure because of lack of
unassigned neighbors.  Such pairs can be taken out of consideration, if,
for instance, their distance exceeds the typical distance between
pairs. The results of vortex and vortex pair detection is illustrated
in Fig.~\ref{fig:pairs}, for a small domain containing 131 vortex pairs.
Most of our production runs contain $\sim 10,000$ pairs at the
beginning of vortex diagnostics.

Post-factum, we have implemented even simpler diagnostics, where the
length of a pair was computed as the distance to closest vortex of the
opposite sign, and obtained qualitatively the same results.

\begin{figure}
\begin{center}
\includegraphics[width=\plotwidth]{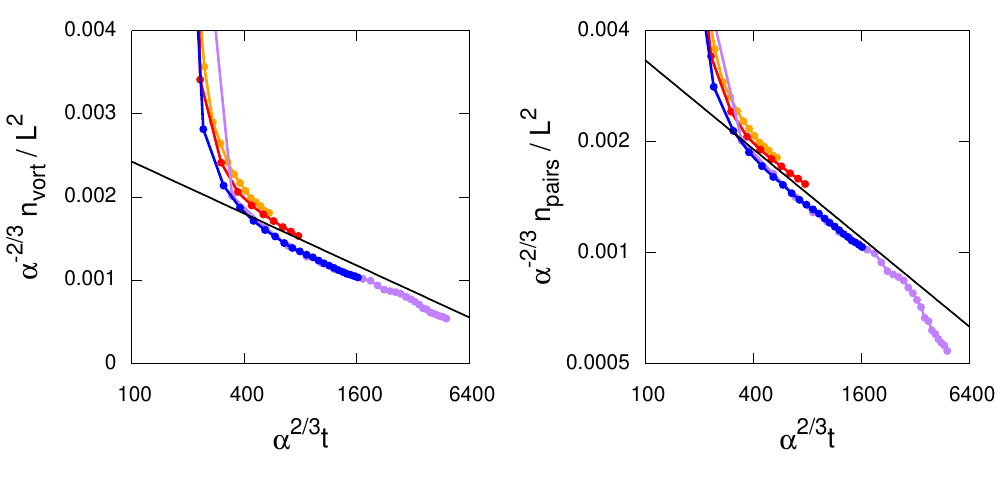}
\caption{
  Number of vortex pairs in lin-log (left) and log-log (right)
  coordinates obtained in simulations shown in
  Fig.~\ref{fig:evol_spectra}, line colors matching.  The longest
  range is for $\alpha = 6400$ in $L=32\pi$ box (blue) and $L=8\pi$
  (purple). The straight lines correspond to
  $f(t) = 0.0045 \, (1-0.1 \ln(\alpha^{2/3}t))$ and to
  $f(t) = 0.021 \, (\alpha^{2/3}t)^{-2/5}$ respectively.
}
\label{fig:nvort}
\end{center}
\end{figure}

The number of vortices, $n_{\rm vort}$, scales with $\alpha$ and
decreases with time, as shown in Fig.~\ref{fig:nvort}. The time range
is too short to distinguish a power law from a logarithmic dependence,
so we restrain from making a statement on the scaling of number of
vortices with time.  Yet, we need $n_{\rm vort}(t)$ dependence to
estimate the probability of small $|\psi|$ and for comparison with
evolution of small systems.  For this purposes, we use the power law
dependence,
\begin{equation}
n_{\rm vort} = 0.021 L^2 \alpha^{2/5} t^{-2/5}.
\label{nvort}
\end{equation}

\begin{figure}
\begin{center}
\includegraphics[width=\plotwidth]{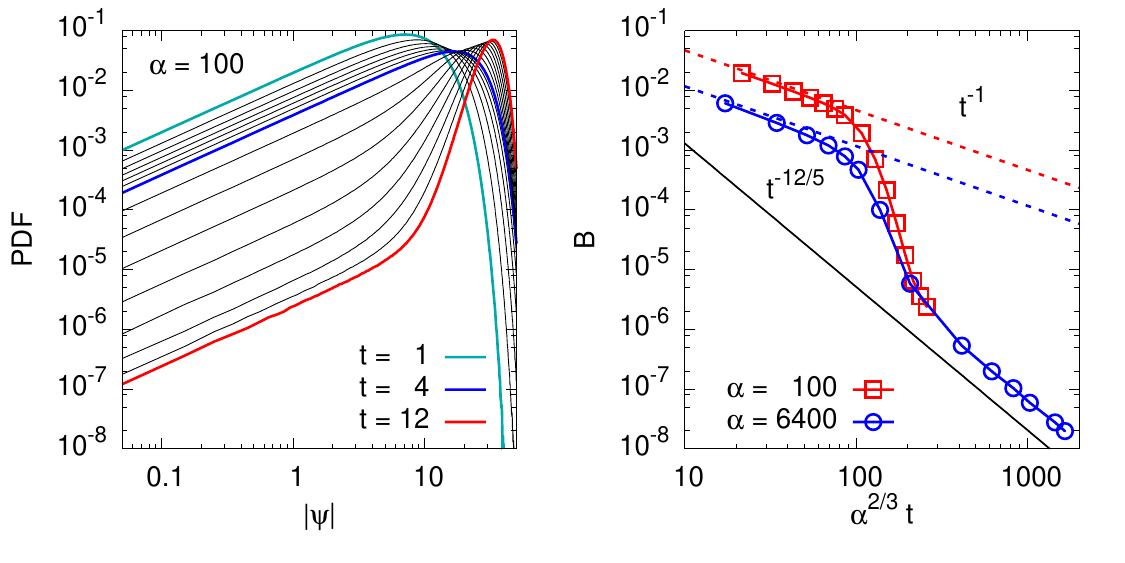}
\caption{
  Left: probability of small $|\psi|$ is a linear function,
  ${\cal P}(|\psi|) = B(t) |\psi|$.  Right: The coefficient $B(t)$ drops at the time
  of transition. At early times, $B = (\tilde{\alpha} t)^{-1}$,
  shown with dashed lines. At later times, $B(t)$ is proportional
  to the vortex density: for the number of vortices given by Eq.~(\ref{nvort}),
  it scales as $B \propto (\alpha^{2/3}t)^{-12/5}$.
}
\label{fig:psipdf_early}
\end{center}
\end{figure}

The proposed scaling explains how the probability of small $|\psi|$
decreases with time.  Initially, probability of small amplitudes is
$2\chi d \chi = 2 N^{-1} |\psi| d |\psi|$, so that ${\cal P}(|\psi|)
\approx B(t)|\psi|$ with $B(t) = 2 N^{-1} = 2 (\tilde{\alpha}
t)^{-1}$.  At later times the probability of small
amplitudes is determined by the density of vortices and by the profile
of individual vortex.  Assuming radially symmetric vortex, one obtains
$|\psi| \sim r$ at the core.  If the healing
length scales as $N^{-1/2}$~\cite{pitaevskii1961}, then $|\psi| \sim N
r$. This leads to $B = 4\pi n_{\rm vort} /(L^2 N^2)$, shown in
Fig.~\ref{fig:psipdf_early}, for $n_{\rm vort}$ given by
Eq.~(\ref{nvort}).  This estimate gives $B \sim (\alpha^{2/3}
t)^{-12/5}$ up to a numerical coefficient.

\begin{figure*}
%\begin{center}
%\includegraphics[width=2.0\plotwidth]{FIGS4/evol4_vortex_histogram.pdf}
\includegraphics[width=2.0\plotwidth]{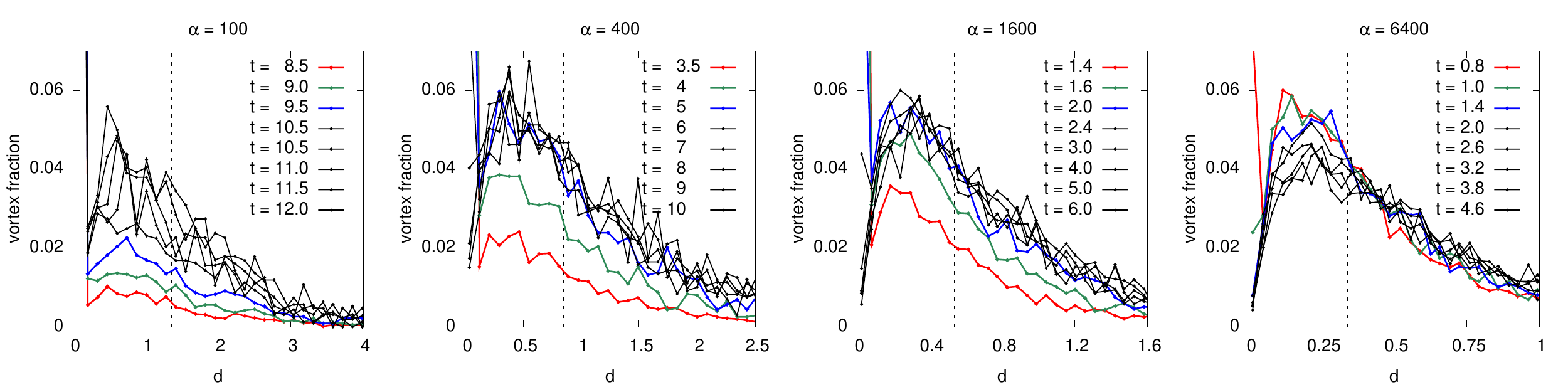}
\caption{
  Histogram of lengths of vortex pairs, obtained in simulations with
  $L=32\pi$.  The plots show the fraction of vortex pairs in a bins
  $(d, d+\Delta d)$ of size $ \Delta d = 0.1 d_2$, where
  $d_2 = 2\pi \alpha^{-1/3}$. The scale $d_2$ is shown with
  dashed vertical lines.  }
\label{fig:dist_hist}
%\end{center}
\end{figure*}

\begin{figure*}
\begin{center}
\includegraphics[width=2.0\plotwidth]{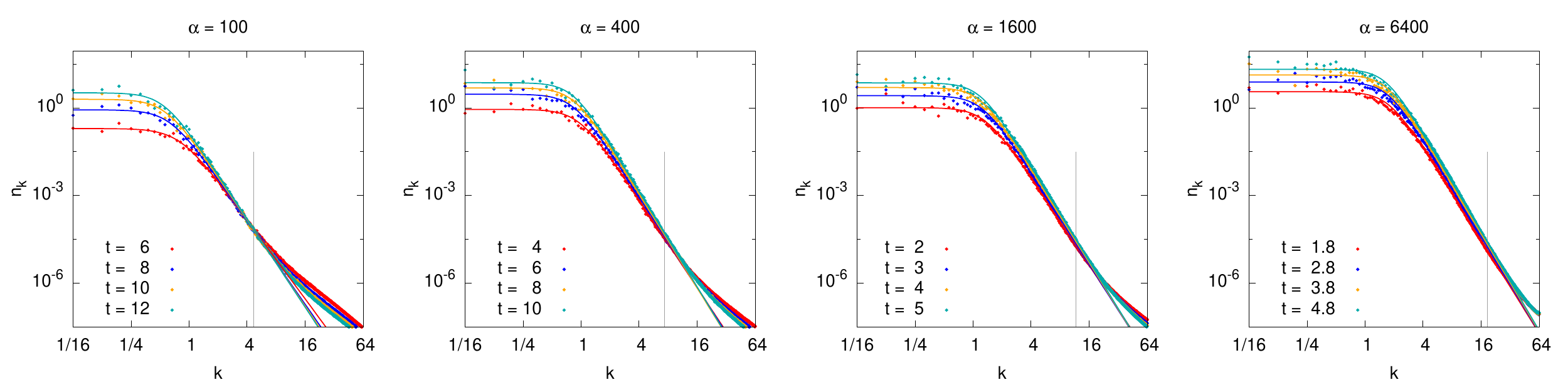}
\caption{
  Spectra at late times for $L=32\pi$ fitted by
  Eq.~\eqref{eq:late_spectra_fit}.  The vertical lines correspond to
  scale $\lambda_2 = 2\pi/k_2$ with $k_2 = \alpha^{1/3}$.
}
\label{fig:late}
\end{center}
\end{figure*}

Both logarithmic and power law scalings for the number of vortices
were reported in literature. Power laws with exponents 0.3-0.4 were
observed in relaxation studies~\cite{schnow2012}, with transitional logarithmic
scalings. Nazarenko and Onorato~\cite{nazono2006} reported a
logarithmic scaling for forced simulations, but the behavior appears
to be transitional as well.  It was observed at the early stages,
before formation of precondensate, with number of vortices in dropping
from 20000 to 3000, in $2\pi$ box, while here the number of pairs is
drops to 2000 in $32\pi$ box.

It is interesting that the straightforward
averaging of inter-vortex distances gives $d_2 \propto d_1 \propto
t^{1/5}$. The contradiction with visual observation of constant $d_2$
is the effects of ultra-short and extra-long dipoles.  During the
stage of vortex formation, vortices are hard to distinguish from
noise; the diagnostics detects colossal number of ``vortex pairs''
with lengths at the limit of resolution.  At later times, isolated
vortices are formally assigned into pairs as a side effect of our
vortex matching algorithm.  The number of such pairs are small, but
their large lengths significantly affect the average. We found the
histograms of the inter-vortex lengths, shown in
Fig.~\ref{fig:dist_hist}, more informative than the average.

In Figure~\ref{fig:dist_hist}, the system with weakest pumping,
$\alpha = 100$ is still going through the vortex formation stage, as
indicated by the peak at the first bin of the distribution. In the case of
$\alpha=6400$, the fraction in the first bin is insignificant for
$t>1$, and the distribution preserves its shape on the course of
system evolution. In all cases, the number of vortices dropped from
$n_{\rm vort} \sim 10000$ to $n_{\rm vort} \sim 1000$ during the
time interval considered. And in all cases, the length $d_2 =
2\pi{\alpha}^{-1/3}$ is proportional to the distance at the the peak of
distribution,  with a \mbox{factor $\sim 2.5$.}

Our observation that the length of a vortex pair depends on the
pumping rate, rather than time, is surprising.  One would expect the
inter-vortex distance to be proportional to a typical size of the
vortex core, which scales as $1/|\psi| \sim
N^{-1/2}$~\cite{pitaevskii1961}.  Such reduction of inter-vortex
distance was observed in experiments~\cite{NeeSam2010} and
simulations~\cite{smimir2012} for vortex pairs moving from regions of
less dense condensate to more dense condensate. In contrast, in images
shown in Fig.~\ref{fig:vortices}, the wave action for system with
$\alpha = 6400$ increases by the factor of 47, which would translate
to the decrease of inter-vortex distances by factor of 7, yet we
observe the inter-vortex distance unchanged.

Nowak at al.~\cite{nowsch2012} made a similar connection between
vortices and the shape of the spectra in simulations on thermalization
of Gross-Pitaevskii turbulence.  First, they inspected the spectrum of
a manufactured field of vortices and concluded that: (i) the spectrum
has $k^{-2}$ slope on the scales greater than the length of a typical
vortex pair; (ii) the slope steepens to $k^{-4}$ for scales between
vortex pair and vortex core; and, (iii) the slope is $k^{-6}$ on
scales below the size of vortex core. Next, they confirmed the
presence of $k^{-2}$ and $k^{-4}$ slopes in dynamical simulations,
(although to observe $k^{-2}$ the authors had to select simulations
with shortest dipoles). As for $k^{-6}$ slope, the interval of smallest
scales was dominated by the spectrum of over-condensate fluctuations,
$k^{-2}$.  Forced evolution has different dynamics than
thermalization.  Even thought both types of spectra show qualitatively
similar shape with three distinct exponents, the values of the exponents
are different.  We observe a plateau (rather than $k^{-2}$ slope) at
largest scales and a mid-range slope that gradually increases with
time.

%-----------------------------------------------------
\subsection{Transition from precondensate to condensate}
\label{supl:condensate}
%-----------------------------------------------------

We expect the transition from precondensate to condensate to occur
when the typical distance between vortex pairs, $d_1$, exceeds the
domain size.  We consider three domain sizes, $L=\pi/2$, $L=\pi$
and $L=2\pi$, and two pumping rates, $\alpha=100$ and $\alpha=1600$,
and we estimate the times of transition to condensate, $t_{\rm cond}$,
as abscissa of $d_1 = L$ in Fig.~\ref{fig:evol_spectra}.  These times
are listed in the Table.  Among considered combinations, the
transition to condensate in cases (a),(b), and (d) is expected to
happen on the border between the stages of vortex formation and vortex
annihilation, $t_{\rm cond} \sim 2 t^*$; for the other three
combinations the transition is expected in the vortex annihilation
regime, $t_{\rm cond} \gg 2 t^*$.  For each case, we have performed 10
simulation with different random seeds. For each realization, we
measure the number of vortex pairs in the domain as function of time.

\begin{table}
\begin{tabular}{ cccccccccccc }
\hline
 &  $\quad\alpha\quad$  & $2 t^*$  & $\quad L \quad$
&  $\;\; L\alpha^{1/3} \;$ & $\;\;\alpha^{-2/3}t_{\rm cond}\;\;$  & $t_{\rm cond}$ \\
\hline
(a) &  100 &  8.4  &  $\pi/2$  &    7.29 &   194  &   9  \\
(b) &  100 &  8.4  &  $\pi$    &   14.58 &   240  &  11 \\
(c) &  100 &  8.4  &  $2\pi$   &   29.16 &  1500  &  70 \\
\hline
(d) & 1600 &  1.3 &  $\pi/2$  &   18.37 &  220 &  1.6 \\
(e) & 1600 &  1.3 &  $\pi$    &   36.74 &  N/A &  $t_{\rm cond}\gg 2t^*$ \\
(f) & 1600 &  1.3 &  $2\pi$   &   73.49 &  N/A &  $t_{\rm cond}\gg 2t^*$ \\
\hline
\end{tabular}
\caption{Parameters of simulations in small boxes and time of
  transition from precondensate to condensate estimated from
  Fig.~\ref{fig:evol_spectra}.  Here, the data beyond interpolation
  range are shown as not available (N/A).}
\end{table}

\begin{figure}
\begin{center}
\includegraphics[width=\plotwidth]{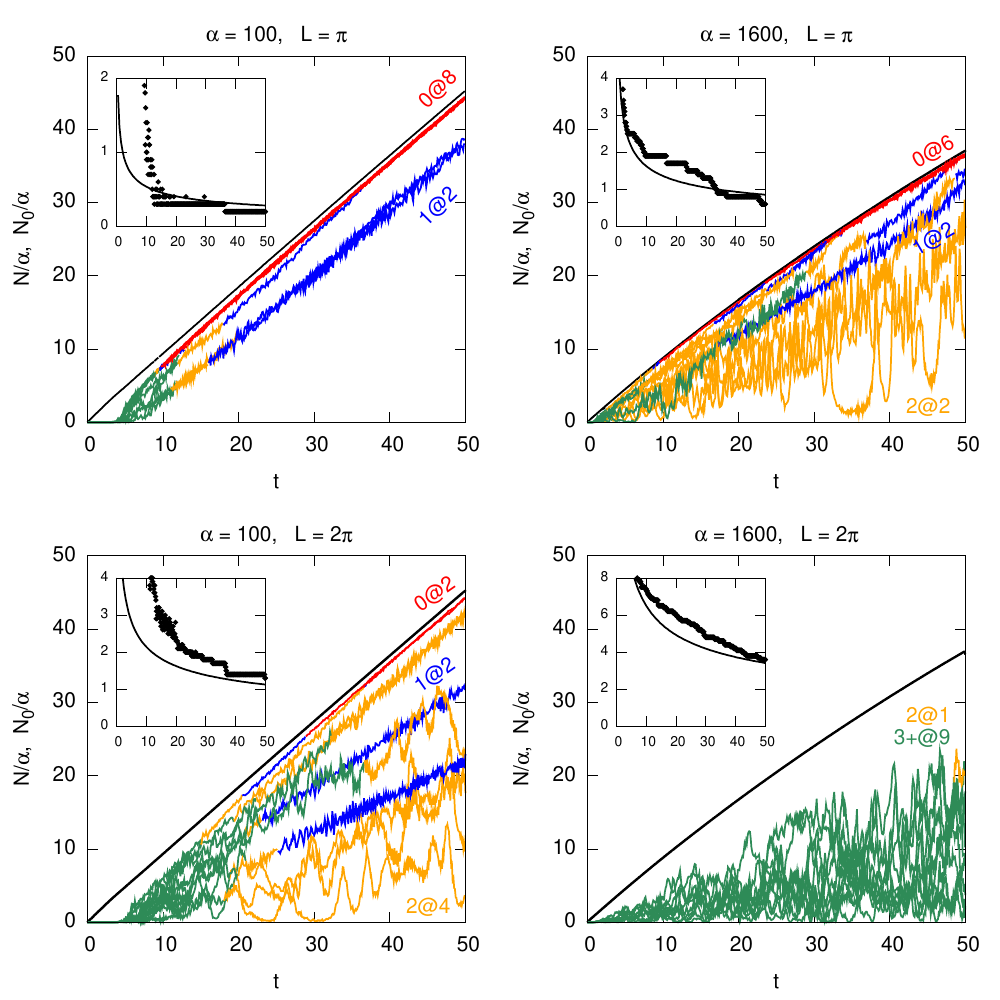}
\caption{
 Establishment of condensate in small domains, $L=\pi$ (above) and
 $L=2\pi$ (below) for $\alpha=100$ (right) and $\alpha=1600$ (left),
 10 realizations per case.  Black lines show the total number of
 waves, $N(t)$, while color lines show the number of waves in the
 condensate, $N_0(t)$.  The number of vortex pairs averaged among
 realizations is shown in insert plots as function of time, next to
 extrapolation obtained from simulations in large domains,
 Eq.~(\ref{nvort}).  The color of $N_0(t)$ curves corresponds to the
 number of vortex pairs in the system: 0 (red), 1 (blue), 2 (yellow),
 3 or more (green). The labels show the state of each ensemble at
 $t=50$ in the format ``$n_{\rm pairs}$@$n_{\rm realizations}$''.
}
\label{fig:cond_N}
\end{center}
\end{figure}

First, let us compare simulations with two different pumping rates in
the domain of size $L=\pi/2$, cases (a) and (d).  In case (a) the last
vortices have disappeared during the time range $[8.0, 10.8]$, in
agreement with expected $t_{\rm cond} = 9$.  In case (d) $t_{\rm
  cond} = 1.6$, and by the time $t=2.6$ six out of ten realizations
are vortex-free. The other four realizations have a single
vortex pair; they become vortex-free by the time $t=16.6$.  This is
consistent with the overall dynamics in large boxes:  at $\alpha=100$
the transition between thermal equilibrium and precondensate regime
occurs relatively late, at $t^*\approx 4.2$, vortices become
detectable at $2t^*\approx 8.4$, and slow vortex annihilation regime
is not reached until $3t^*\approx 12.6$.  On the other hand, for
$\alpha=1600$, $2t^*\approx 1.3$ which explains disappearance of most
of the vortices by time $t=2.6$. Vortex annihilation becomes slow
after $3t^* \approx 2$, that's why it takes so long time (up to
$t=16.6$) for remaining four pairs to disappear.

Simulations with $L=\pi$ and $L=2\pi$, show qualitatively the same
results, except that only the runs from case (b) have good chances of
forming system-wide condensates before vortex annihilation becomes
slow.

For cases (b),(c),(e), and (f) we compare the wave action $N_0(t)$
accumulated in the condensate (that is in the $k=0$ mode) to the wave
action of whole system, $N(t)$. The comparison is shown in
Fig.~\ref{fig:cond_N}.  Within each set of realizations, the curve
$N(t)$ does not depend on realization.  (The curves for $\alpha=1600$
deviate from linear growth because of higher losses to damping at
large $N$.)  In contrast, the wave action in the condensate is
different in each realization, at least during the time when vortices
are still present in the system.  When vortices are gone, the wave
action of over-condensate fluctuations, $N-N_0$, stays at
an approximately constant level, with the exceptions of small-amplitude
oscillations \cite{milvla2013}.  Notice that the small level of
over-condensate fluctuations, $N-N_0\ll N$, does not guarantee a
vortex-free system.  Moreover, the systems with the same number of
vortices can have different fractions of waves in the condensate, and
$N - N_0$ is non-monotonic function of the number of vortex pairs.

The inserts in Fig.~\ref{fig:cond_N} provide another way to compare
the dynamics of vortex annihilation in small and large domains.  Here,
the dots show the number of vortex pairs in small domains, averaged
over 10 realization, as a function of time.  The lines are predictions
derived from extrapolation, Eq.~\eqref{nvort}, for large domains.
Qualitatively, the number of vortex pairs in small systems agree with
dynamics of evolution of large systems.

%-----------------------------------------------------
\subsection{Timescale of transition in physical units}
\label{supl:units}
%-----------------------------------------------------

The timescale $t^*$ is an important characteristic of the system.  Our
simulations, done in non-dimensional variables, show that $\tilde{t}^*
\approx 90 ( d\tilde{N}/d\tilde{t})^{-2/3}$. (In this Supplement we
denote non-dimensionalized quantities by tildes.)  Let us estimate $t^*$
for a physical system.

We restore physical
dimensions in Eq.~\eqref{NLSE},
\[
i q^2 \tau \psi_t + q^2 \ell^2 \nabla^2 \psi \pm q^2 \frac{|\psi|^2}{I_0} \psi = 0,
\]
introducing coefficients $\tau$, $\ell$, and $I_0$ that have units of
time, length, and wave intensity respectively.  The multiplier $q$ is an
arbitrary quantity that parametrizes the family of transformations
between simulation units and physical units,
\[
t = q^2 \tau \tilde{t}, \qquad
x =  q \ell \tilde{x}, \qquad
\psi = \frac{\sqrt{I_0}}{q} \tilde{\psi}.
%,\qquad N = \frac{I_0}{q^2} \tilde{N}.
\]
It is natural to assume that the physical pumping scale, $\ell_p$, is
known. Then, we can use it to select transformation parameter, $q =
\ell_p / (\ell \tilde{\ell}_p )$, where $\tilde{\ell_p} = 2\pi/80$ is
the pumping scale in our simulation units. Thus, we obtain,
\[
   t^* \approx 16.5 \, \tau \left[\frac{\ell_p}{\ell} \frac{\tau}{I_0} \dot{N} \right]^{-2/3},
\]
where $N = \langle|\psi|^2\rangle$.

%-----------------------------------------------------

%\section*{To do}

%See if any of these articles are relevant,
%\cite{connew2003,conjos2005,durpic2009,jospom2006,nazono2006,pronaz2009,%
%pic2007,picric2008,picben2009,picgar2014}.

%-----------------------------------------------------

%G.F. work is supported by
%grants of Bi-National Science Foundation, Minerva Foundation with
%funding from the German Ministry for Education and Research and by
%Russian Science Foundation project No. 14-22-00259 (development of the
%analytical theory and writing the paper).

%------------------------------------------------------------------------

\bibliography{cascade,biblionls}

\begin{thebibliography}{17}
\expandafter\ifx\csname natexlab\endcsname\relax\def\natexlab#1{#1}\fi
\expandafter\ifx\csname bibnamefont\endcsname\relax
  \def\bibnamefont#1{#1}\fi
\expandafter\ifx\csname bibfnamefont\endcsname\relax
  \def\bibfnamefont#1{#1}\fi
\expandafter\ifx\csname citenamefont\endcsname\relax
  \def\citenamefont#1{#1}\fi
\expandafter\ifx\csname url\endcsname\relax
  \def\url#1{\texttt{#1}}\fi
\expandafter\ifx\csname urlprefix\endcsname\relax\def\urlprefix{URL }\fi
\providecommand{\bibinfo}[2]{#2}
\providecommand{\eprint}[2][]{\url{#2}}

\bibitem[{\citenamefont{Zakharov et~al.}(1992)\citenamefont{Zakharov, Lvov, and
  Falkovich}}]{ZakharovLvovFalkovich1992}
\bibinfo{author}{\bibfnamefont{V.~E.} \bibnamefont{Zakharov}},
  \bibinfo{author}{\bibfnamefont{V.~S.} \bibnamefont{Lvov}}, \bibnamefont{and}
  \bibinfo{author}{\bibfnamefont{G.}~\bibnamefont{Falkovich}},
  \emph{\bibinfo{title}{Kolmogorov Spectra of Turbulence I: Wave turbulence}}
  (\bibinfo{publisher}{Springer-Verlag}, \bibinfo{address}{New York},
  \bibinfo{year}{1992}).

\bibitem[{\citenamefont{Falkovich and Shafarenko}(1991)}]{falsha1991}
\bibinfo{author}{\bibfnamefont{G.}~\bibnamefont{Falkovich}} \bibnamefont{and}
  \bibinfo{author}{\bibfnamefont{A.}~\bibnamefont{Shafarenko}},
  \bibinfo{journal}{Journal of Nonlinear Science} \textbf{\bibinfo{volume}{1}},
  \bibinfo{pages}{457} (\bibinfo{year}{1991}).

\bibitem[{\citenamefont{Connaughton et~al.}(2003)\citenamefont{Connaughton,
  Newell, and Pomeau}}]{connew2003}
\bibinfo{author}{\bibfnamefont{C.}~\bibnamefont{Connaughton}},
  \bibinfo{author}{\bibfnamefont{A.~C.} \bibnamefont{Newell}},
  \bibnamefont{and} \bibinfo{author}{\bibfnamefont{Y.}~\bibnamefont{Pomeau}},
  \bibinfo{journal}{Physica D: Nonlinear Phenomena}
  \textbf{\bibinfo{volume}{184}}, \bibinfo{pages}{64} (\bibinfo{year}{2003}).

\bibitem[{\citenamefont{Pitaevskii and
  Stringari}(2003)}]{PitaevskiiStringariBook2003}
\bibinfo{author}{\bibfnamefont{L.~P.} \bibnamefont{Pitaevskii}}
  \bibnamefont{and}
  \bibinfo{author}{\bibfnamefont{S.}~\bibnamefont{Stringari}},
  \emph{\bibinfo{title}{Bose-Einstein Condensation}}
  (\bibinfo{publisher}{Clarendon}, \bibinfo{address}{Oxford},
  \bibinfo{year}{2003}).

\bibitem[{\citenamefont{Dyachenko et~al.}(1992)\citenamefont{Dyachenko, Newell,
  Pushkarev, and Zakharov}}]{dnpz}
\bibinfo{author}{\bibfnamefont{S.}~\bibnamefont{Dyachenko}},
  \bibinfo{author}{\bibfnamefont{A.~C.} \bibnamefont{Newell}},
  \bibinfo{author}{\bibfnamefont{A.}~\bibnamefont{Pushkarev}},
  \bibnamefont{and} \bibinfo{author}{\bibfnamefont{V.~E.}
  \bibnamefont{Zakharov}}, \bibinfo{journal}{Physica D}
  \textbf{\bibinfo{volume}{57}}, \bibinfo{pages}{96} (\bibinfo{year}{1992}).

\bibitem[{\citenamefont{Nazarenko and Onorato}(2006)}]{nazono2006}
\bibinfo{author}{\bibfnamefont{S.}~\bibnamefont{Nazarenko}} \bibnamefont{and}
  \bibinfo{author}{\bibfnamefont{M.}~\bibnamefont{Onorato}},
  \bibinfo{journal}{Physica D: Nonlinear Phenomena}
  \textbf{\bibinfo{volume}{219}}, \bibinfo{pages}{1} (\bibinfo{year}{2006}).

\bibitem[{\citenamefont{Sulem and Sulem}(1999)}]{SulemSulem1999}
\bibinfo{author}{\bibfnamefont{C.}~\bibnamefont{Sulem}} \bibnamefont{and}
  \bibinfo{author}{\bibfnamefont{P.~L.} \bibnamefont{Sulem}},
  \emph{\bibinfo{title}{Nonlinear Schr\"odinger Equations: Self-Focusing and
  Wave Collapse}} (\bibinfo{publisher}{World Scientific}, \bibinfo{address}{New
  York}, \bibinfo{year}{1999}).

\bibitem[{\citenamefont{Bogoliubov}(1947)}]{Bogoliubov1947}
\bibinfo{author}{\bibfnamefont{N.}~\bibnamefont{Bogoliubov}},
  \bibinfo{journal}{J. Phys. (USSR)} \textbf{\bibinfo{volume}{11}},
  \bibinfo{pages}{23} (\bibinfo{year}{1947}).

\bibitem[{\citenamefont{Dyachenko and
  Falkovich}(1996)}]{DyachenkoFalkovich1996}
\bibinfo{author}{\bibfnamefont{A.}~\bibnamefont{Dyachenko}} \bibnamefont{and}
  \bibinfo{author}{\bibfnamefont{G.}~\bibnamefont{Falkovich}},
  \bibinfo{journal}{Phys. Rev. E} \textbf{\bibinfo{volume}{54}},
  \bibinfo{pages}{5095} (\bibinfo{year}{1996}).

\bibitem[{\citenamefont{Nowak et~al.}(2012)\citenamefont{Nowak, Schole, Sexty,
  and Gasenzer}}]{nowsch2012}
\bibinfo{author}{\bibfnamefont{B.}~\bibnamefont{Nowak}},
  \bibinfo{author}{\bibfnamefont{J.}~\bibnamefont{Schole}},
  \bibinfo{author}{\bibfnamefont{D.}~\bibnamefont{Sexty}}, \bibnamefont{and}
  \bibinfo{author}{\bibfnamefont{T.}~\bibnamefont{Gasenzer}},
  \bibinfo{journal}{Physical Review A} \textbf{\bibinfo{volume}{85}},
  \bibinfo{pages}{043627} (\bibinfo{year}{2012}).

\bibitem[{\citenamefont{Schole et~al.}(2012)\citenamefont{Schole, Nowak, and
  Gasenzer}}]{schnow2012}
\bibinfo{author}{\bibfnamefont{J.}~\bibnamefont{Schole}},
  \bibinfo{author}{\bibfnamefont{B.}~\bibnamefont{Nowak}}, \bibnamefont{and}
  \bibinfo{author}{\bibfnamefont{T.}~\bibnamefont{Gasenzer}},
  \bibinfo{journal}{Physical Review A} \textbf{\bibinfo{volume}{86}},
  \bibinfo{pages}{013624} (\bibinfo{year}{2012}).

\bibitem[{\citenamefont{Falkovich and Vladimirova}(2015)}]{falvla2015}
\bibinfo{author}{\bibfnamefont{G.}~\bibnamefont{Falkovich}} \bibnamefont{and}
  \bibinfo{author}{\bibfnamefont{N.}~\bibnamefont{Vladimirova}},
  \bibinfo{journal}{Phys. Rev. E} \textbf{\bibinfo{volume}{91}},
  \bibinfo{pages}{041201} (\bibinfo{year}{2015}).

\bibitem[{\citenamefont{Vladimirova et~al.}(2012)\citenamefont{Vladimirova,
  Derevyanko, and Falkovich}}]{vlader2012}
\bibinfo{author}{\bibfnamefont{N.}~\bibnamefont{Vladimirova}},
  \bibinfo{author}{\bibfnamefont{S.}~\bibnamefont{Derevyanko}},
  \bibnamefont{and}
  \bibinfo{author}{\bibfnamefont{G.}~\bibnamefont{Falkovich}},
  \bibinfo{journal}{Physical Review E} \textbf{\bibinfo{volume}{85}},
  \bibinfo{pages}{010101} (\bibinfo{year}{2012}).

\bibitem[{\citenamefont{Pitaevskii}(1961)}]{pitaevskii1961}
\bibinfo{author}{\bibfnamefont{L.}~\bibnamefont{Pitaevskii}},
  \bibinfo{journal}{Sov. Phys. JETP} \textbf{\bibinfo{volume}{13}},
  \bibinfo{pages}{451} (\bibinfo{year}{1961}).

\bibitem[{\citenamefont{Neely et~al.}(2010)\citenamefont{Neely, Samson,
  Bradley, Davis, and Anderson}}]{NeeSam2010}
\bibinfo{author}{\bibfnamefont{T.~W.} \bibnamefont{Neely}},
  \bibinfo{author}{\bibfnamefont{E.~C.} \bibnamefont{Samson}},
  \bibinfo{author}{\bibfnamefont{A.~S.} \bibnamefont{Bradley}},
  \bibinfo{author}{\bibfnamefont{M.~J.} \bibnamefont{Davis}}, \bibnamefont{and}
  \bibinfo{author}{\bibfnamefont{B.~P.} \bibnamefont{Anderson}},
  \bibinfo{journal}{Phys. Rev. Lett.} \textbf{\bibinfo{volume}{104}},
  \bibinfo{pages}{160401} (\bibinfo{year}{2010}),
  \urlprefix\url{https://link.aps.org/doi/10.1103/PhysRevLett.104.160401}.

\bibitem[{\citenamefont{Smirnov and Mironov}(2012)}]{smimir2012}
\bibinfo{author}{\bibfnamefont{L.}~\bibnamefont{Smirnov}} \bibnamefont{and}
  \bibinfo{author}{\bibfnamefont{V.}~\bibnamefont{Mironov}},
  \bibinfo{journal}{Physical Review A} \textbf{\bibinfo{volume}{85}},
  \bibinfo{pages}{053620} (\bibinfo{year}{2012}).

\bibitem[{\citenamefont{Miller et~al.}(2013)\citenamefont{Miller, Vladimirova,
  and Falkovich}}]{milvla2013}
\bibinfo{author}{\bibfnamefont{P.}~\bibnamefont{Miller}},
  \bibinfo{author}{\bibfnamefont{N.}~\bibnamefont{Vladimirova}},
  \bibnamefont{and}
  \bibinfo{author}{\bibfnamefont{G.}~\bibnamefont{Falkovich}},
  \bibinfo{journal}{Phys. Rev. E} \textbf{\bibinfo{volume}{87}},
  \bibinfo{pages}{065202} (\bibinfo{year}{2013}).

\end{thebibliography}

%\include{cascade10a}

%------------------------------------------------------------------------
\end{document}